\documentclass{eas}
\usepackage{graphicx}
\usepackage{xcolor}        
\usepackage{enumitem} 
\usepackage{natbib}
\usepackage{array}

\begin{document}

\title{Automated eclipsing binary detection:\\applying the Gaia CU7 pipeline to Hipparcos} 
\runningtitle{Holl \etal: Hipparcos eclipsing binary detection with Gaia CU7 pipeline}

\author{Berry Holl}\address{Department of Astronomy, University of Geneva, CH 1290 Versoix  (\email{berry.holl@unige.ch})}
\author{Nami Mowlavi}\sameaddress{1}
\author{Isabelle Lecoeur-Ta\"ibi}\sameaddress{1}
\author{Fabio Barblan}\sameaddress{1}
\author{Lorenzo Rimoldini}\sameaddress{1}
\author{Laurent Eyer}\sameaddress{1}
\author{Maria S\"uveges}\sameaddress{1}
\author{Leanne Guy}\sameaddress{1}
\author{Diego Ordo\~nez-Blanco}\sameaddress{1}
\author{Idoia Ruiz}\sameaddress{1}
\author{Krzysztof Nienartowicz}\sameaddress{1}

\begin{abstract}
\vspace{-4mm}
We demonstrate the eclipsing binary detection performance of the Gaia variability analysis and processing pipeline using Hipparcos 
data. The automated pipeline classifies 1\,067 (0.9\%) of the 118\,204 Hipparcos sources as eclipsing binary candidates.
The detection rate amounts to 89\% (732 sources) in a subset of 819 visually confirmed eclipsing binaries, with the period correctly identified for 80\% of them, and double or half periods obtained in 6\% of the cases.
\end{abstract}

\maketitle

\vspace{-4mm}
\section{Introduction}
\vspace{-3mm}
The Gaia mission is expected to observe $\sim$1~billion sources, among which 0.4 to 7~million are expected to be eclipsing binaries \citep[as summarized in][]{2013EAS....64..399H}.
The detection and characterisation of those eclipsing binaries are distributed over two Coordination Units (CUs) of the Gaia Data Processing and Analysis Consortium.
CU7 has the task to identify eclipsing binaries, find their orbital periods, and sub-classify them.
This information is then passed to CU4, who will model the eclipsing binaries and derive stellar and orbital parameters \citep{2012ocpd.conf...59S}.
The processing loop is schematized in Fig.~1 of \cite{2013EAS....64..399H}.

In this short contribution, we demonstrate the eclipsing binary detection performance of the current Gaia CU7 pipeline
on the Hipparcos data \citep{1997ESASP1200.....E}. Because the Hipparcos time sampling and mean number of 
observations are similar to Gaia, it a good dataset to test the performance of the Gaia processing pipeline.

The Hipparcos data set is described in Sect.~\ref{sect:referenceSet} and the results are presented in Sect.~\ref{sect:results}.
Conclusions are drawn in Sect.~\ref{sect:conclusions}.


\section{Hipparcos eclipsing binaries}\label{sect:referenceSet}
\vspace{-3mm}
We define a reference set of 819 Hipparcos eclipsing binaries after visual inspection of the light curves, chosen among the list of eclipsing binaries published in Vol.~11 of the ``Hipparcos catalogue of periodic variable stars'' of \cite{1997ESASP1200.....E}, to which additional Hipparcos sources are added that are flagged as eclipsing binaries in the January 2014 revision of the AAVSO catalogue \citep{Watson13}.
We must note that the periods published in the Hipparcos catalog were derived from Hipparcos light curves for only 682 sources.
The periods published for the remaining 137 sources were taken from the literature.
We thus do not expect our automated pipeline to recover the periods of all those later sources.


\section{Gaia CU7 pipeline applied to Hipparcos}\label{sect:results}
\vspace{-3mm}
The Gaia CU7 pipeline, outlined in \cite{Eyer15}, can be divided in four per-source analysis steps.
They are briefly described in the following sections, with our application to the 115\,423 Hipparcos sources that have at least one good observation (flag 0 or 1) in their time series. Note that the applied pipeline configuration is simplified with respect to what is planned for official (Gaia) data processing.

\begin{table}[t]
\centering
\begin{tabular}{l|rl|rl}
\hline 
Module & \multicolumn{2}{c|}{All Hipparcos} & \multicolumn{2}{c}{Eclipsing binaries ref. set} \\
 & \#sources & \phantom{100.00}\% & \#sources & \phantom{100.0}\% \\
\hline
Input time series & 115\,423 & 100.00\% & 819 & 100.0\% \\
Variability detection & 15\,568 & \phantom{1}13.49\% & 819 & 100.0\% \\
Supervised clas.: eclipsing & 1\,598 & \phantom{10}1.38\% & 752 & \phantom{1}91.8\% \\
SOS Eclipsing binaries & 1\,067 & \phantom{10}0.92\% & 732 & \phantom{1}89.4\% \\
\hline
\end{tabular}
\caption{Pipeline processing result from top to bottom for the whole Hipparcos catalog (left), and for the eclipsing binaries reference set~(right).}
\label{tab:resultSelectivity}
\end{table}

\vspace{-3mm}
\subsection{Variability detection}
\vspace{-2mm}
Variable sources are detected using a $\chi^2$ test with a p-value threshold of $p<10^{-4}$.
This reduces the list of time series to be processed to 15\,568, see Table~\ref{tab:resultSelectivity}.

\vspace{-3mm}
\subsection{Characterisation}
\vspace{-2mm}
Periodic sources are searched using the unweighted Lomb-Scargle method \citep{1976Ap&SS..39..447L,1982ApJ...263..835S}, and multi-harmonic Fourier series are fitted to their light curves.


\vspace{-3mm}
\subsection{Classification}
\vspace{-2mm}
The 15,568 variable sources are classified into 23 different variable types using a supervised Random Forest classifier \citep{Breiman01}. Input to the classifier are various Hipparcos specific attributes \citep[detailed in][]{2011MNRAS.414.2602D} which are derived from the light curve model parameters determined in the previous step, together with the parallax, and V-I colour.
Eclipsing binaries are represented by one class containing types `EA', `EB', and `EW'.
We base our training set on \cite{2011MNRAS.414.2602D}, which however includes 72\% of the eclipsing binaries in our reference data set (Sect.~\ref{sect:referenceSet}). To make the reference set more independent of the training set, we train our classifier with only half of the Dubath training set, containing only 37\% of the reference eclipsing binaries. Although this reduces the precision of the classifier, it strengthens the power of the reference set to evaluate the pipeline performance.
The (ten-fold cross-validation) confusion matrix of the training set has a high completeness\footnote{Type completeness\,\,\, = number of correctly classified sources / number in training set.} of 94.7\% and a contamination\footnote{Type contamination = 1 - number of correctly classified sources / number classified.} of only 7.5\% for eclipsing binaries.

Table~\ref{tab:resultSelectivity} shows that, applied to all Hipparcos data, the trained Random Forest classifier predicts 1\,598 sources to be of type eclipsing binary (selecting those with probability $>0.5$), which includes 92\% of the reference eclipsing binaries. 


\begin{figure}[t]
\centering
\includegraphics[width=\textwidth]{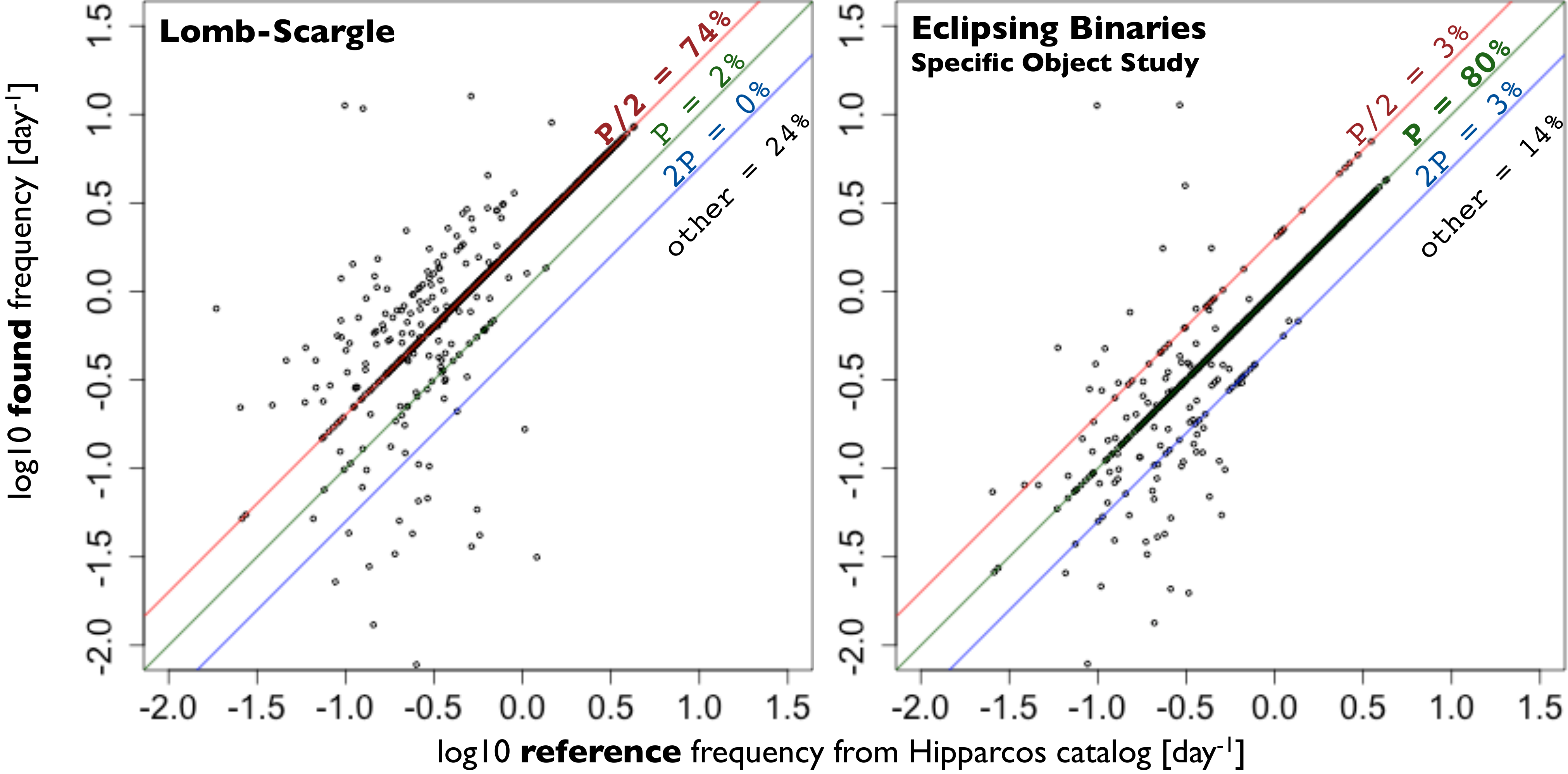}
\caption{Period recovery for the 732 eclipsing binaries from the reference set that were identified as eclipsing binaries.
The results from the unweighted Lomb-Scargle period search (left), and the `corrected' periods using our 
Specific Object Study module `Eclipsing Binaries' (right) recovering the correct period in most cases. }
\label{fig:periodRecovery}
\end{figure}
 
\vspace{-2mm}
\subsection{Specific Object Studies (SOS): Eclipsing Binaries}
\vspace{-2mm}
This post-classification step processes all 1\,598 eclipsing binary candidates provided by supervised classification.
It aims at finding the best period and at sub-classifying the eclipsing binaries based on a two-Gaussian model description of the eclipses in the folded light curves. The detection and period identification algorithms will be detailed in \cite{HollInPrep}, and the two-Gaussian modeling and sub-classification algorithm in \cite{MowlaviInPrep}.
Basically, the automated procedures test the goodness of the model fits for several fractions and multiples of the computed Lomb Scargle period.
If not satisfactory, the procedure is repeated for additional periods found with phase-dispersion minimisation \citep{ Jurkevich71, Stellingwerf78, 1997ApJ...489..941S} and String Length \citep{Lafler65, Burke70}.
The best period is retained from those tests.
The eclipsing binary is then sub-classified based on the two-Gaussian model parameters as described in \cite{MowlaviInPrep}.

This last step of per-source pipeline processing confirms 1\,067 eclipsing binaries (Table~\ref{tab:resultSelectivity}), of which 89\% (732) are in our reference set of eclipsing binaries.
Figure~\ref{fig:periodRecovery} shows that of these 732 reference eclipsing binaries, SOS Eclipsing Binaries recovers\footnote{$\hbox{The period X*P}_{true}\hbox{ is recovered if } \left| \hbox{P}_{true} - \hbox{P}_{found}/X \right| \left(\Delta T /\hbox{P}_{true}\right)<0.1\hbox{P}_{true}$, with $\Delta T$ the time-series duration and $\hbox{P}_{true}$ the Hipparcos or literature period, see \cite{2011MNRAS.414.2602D}.} the 
periods listed in the Hipparcos catalog for 80\% of the sources, and double or half periods for 6\%. As mentioned in Sect.~\ref{sect:referenceSet}, the periods in the Hipparcos catalog were taken from literature for 137 sources. The SOS Eclipsing Binaries package identifies 113 of those and recovers the period for 48\%, and  double or half the period for 7\%.

\vspace{-2mm}
\section{Conclusions and discussion}\label{sect:conclusions}
\vspace{-3mm}
Applying the Gaia CU7 pipeline to the 118\,204 Hipparcos time series, 1\,067~(0.9\%) are identified as eclipsing binary candidates.
Validating the results against a reference set of 819 visually identified eclipsing binaries we find that 732 (89\%) are included, for which the Hipparcos period is recovered in 80\% of the cases. Assuming that the reference set contains all eclipsing binaries detectable in Hipparcos data translates into a completeness of 89\% and contamination of 31\% of our 1\,067 candidates. 
Further investigation of this 31\% `contamination' is planned to be included in \cite{HollInPrep}. 

Given the similarities between Hipparcos and Gaia observations, the good detection rate on Hipparcos data suggests that the current CU7 variability analysis and processing pipeline is in good shape to automatically detect eclipsing binaries in Gaia too.


\end{document}